\documentclass{aastex}
\slugcomment{to be published in the  Astrophysical Journal, Vol. 625, May 
20, 2005}
\lefthead{Geballe et al.}
\righthead{U Equulei's Circumstellar Gas}

\begin{document}

\title{Infrared Spectroscopy of U Equulei's Warm Circumstellar
Gas}

\author{T. R. Geballe\altaffilmark{1}, C. Barnbaum\altaffilmark{2}, Keith
S. Noll\altaffilmark{3}, and M. Morris\altaffilmark{4}} 
\altaffiltext{1}{Gemini Observatory, 670 N. A'ohoku Place, Hilo, HI
96720; tgeballe@gemini.edu}
\altaffiltext{2}{Department of Physics, Astronomy, \& Geosciences,
Valdosta State University; Valdosta, GA 31698; cbarnbaum@valdosta.edu}
\altaffiltext{3}{Space Telescope Science Institute, 3700 San 
Martin Dr., Baltimore, MD 21218; noll@stsci.edu}
\altaffiltext{4}{Department of Physics and Astronomy, University of
California at Los Angeles, Los Angeles, CA 90095; morris@astro.ucla.edu}

\begin{abstract}

Medium and high resolution spectroscopy of U Equulei from 1 to 4 $\mu$m
during 1997-2003 has revealed information about its unusual circumstellar
envelope, observed previously at optical and radio wavelengths. Strong
absorption bands of H$_{2}$O and of CO dominate the 1--4 $\mu$m spectrum.
The gas has a mean temperature of 600~K and $^{12}$C/$^{13}$C$\leq$10. The
CO 2-0 line profiles and velocities imply no net ejection or infall and
indicate either rapid radial gas motions being seen along a narrow
continuum beam, or absorption by orbiting gas that is nearly coincident
with a highly extended continuum source. The gas could be located in a
disk-like structure.  The observed high column densities of warm CO and
H$_{2}$ normally would be associated with sufficient dust to completely
obscure the star at optical wavelengths. The observations thus indicate
either a highly abnormal gas-to-dust ratio, consistent with the earlier
optical observation of abundant refractory metal oxides in the
circumstellar gas, or peculiar geometry and/or illumination.

\end {abstract}


\section{Introduction}

The peculiar variable star, U Equulei (IRAS~20547+0247), has been
described by Barnbaum et al. (1996) as having one of the most
unusual optical spectra ever observed. Their spectra, obtained in 1994,
are dominated by strong absorption bands of metallic oxides, superficially
similar to those of mid-late M giants. However the TiO, AlO and VO in U
Equ are simultaneously in absorption and in emission.  In addition, the
bands of these molecules are finely structured, implying a circumstellar,
rather than photospheric origin for them.  The absorption features must
arise from a region so cool that the higher rotational levels are not
populated. Yet, at such a temperature the titanium seen in TiO should be
in grains.

Besides its unique optical spectrum U Equ displays other peculiarities.
Its OH and H$_{2}$O maser emission (Sivagnanam et al. 1990 and Zuckerman
\& Lo 1987, respectively) vary in velocity as well in strength. U Equ has
a large radial velocity of -78~km~s$^{-1}$ (LSR, -91 ~km~s$^{-1}$
heliocentric; Sivagnanam et al. 1990); the uncertainty in this value may
be as much as 10--20~km~s$^{-1}$ (C. Barnbaum 1996, unpublished data). The
object is also at high galactic latitude (--26$^\circ$), suggesting a
possible link to a halo population. The 25/12~$\mu$m color observed by the
Infrared Astronomical Satellite (IRAS) is typical of optically thin
circumstellar envelopes, despite its IRAS LRS spectrum displaying a
moderately strong 10~$\mu$m silicate absorption band (Barnbaum et al.
1996, Chen \& Gao 2002), indicating a dusty envelope. Almost all other
oxygen-rich stars with similar 25/12 $\mu$m colors show the 10 $\mu$m
feature in emission, consistent with optically thin shells.

Barnbaum et al. (1996) suggested that U Equ's silicate absorption feature
arises in a thick, dusty circumstellar disk.  Since the 25/12 $\mu$m
colors indicate optically thin dust, they proposed that the dusty disk is
seen edge-on, with an extended spherical or asymmetrical component giving
rise to the 12 and 25 $\mu$m emission.

U Equ is listed in the General Catalogue of Variable Stars as an irregular
long period variable (Lb) with a magnitude variation in the P
(photographic) band of 14.5 to 15.5.  Barnbaum et al. (1996) estimate its
distance as 1.5~kpc. In 1994, the visual magnitude was approximately 9,
yet when it was observed again in 1996, 1998, and 1999, the star had faded
to V=13. The variations in the the visual magnitude and the origin of the
circumstellar material are unexplained.  However, the unusual nature of
the optical spectrum suggests that U Equ might have entered a rapid stage
of evolution. Siess \& Livio (1999) have considered that accretion of a
giant planet onto a growing red giant could spin up the stellar envelope
sufficiently to extrude an equatorially compressed outflow, or expanding
disk, and that this might resemble the peculiar environment of U Equ.

In this paper we report infrared spectroscopy of U Equ, which allows a
fuller characterization of the physical conditions in the circumstellar
material. In particular and in corroboration of the optical study of
Barnbaum et al. (1996), we also have found a large quantity of absorbing
gaseous matter that is difficult to reconcile with the relatively modest
amount of extinction by dust.

\section{Observations}

Spectra of U Equ in the 1.0--2.5 and 2.9--4.1~$\mu$m regions were obtained
at the 3.8~m diameter United Kingdon Infrared Telescope (UKIRT) at various
times during 1997--2003.  An observing log is provided in Table~1. The
medium resolution observations employed UKIRT's facility spectrographs:
CGS4 (Mountain et al. 1990) with its 40 and 150 l/mm gratings, and
0.6$\arcsec$ wide slit; and UIST (Ramsay Howat et al. 2000) with its HK
grism and 0.48$\arcsec$ wide slit. Resolving powers ranged from 450 to
4100. The spectra were obtained in the stare~-~nod-along-slit mode. The
calibration stars listed in the table were observed immediately before or
after U Equ for the purpose of flux calibration and removal of telluric
absorption lines. Fluxes of stars in the various photometric bands were
estimated from their visual magnitudes and visible-infrared colors
predicted for their spectral types (Tokunaga 2000).  Wavelength
calibration was derived from spectra of arc lamps and is accurate to
considerably better than 0.001~$\mu$m. The echelle in CGS4 was also used
in 2003 to obtain high resolution spectra in a number of narrow wavelength
intervals in the K window at resolving powers of about 20,000,
corresponding to a velocity resolution of 15~km~s$^{-1}$). For these
spectra wavelength calibrations were obtained from telluric absorption
lines and are accurate to 3~km~s$^{-1}$.

Data reduction employed Figaro routines to extract source spectra from the
spectral images produced by the CGS4 and UIST arrays, wavelength-calibrate
them, ratio them, and perform an initial flux calibration. It was found
that where the lower resolution $JHK$ and $LL'$ spectral segments
overlapped or adjoined one another flux levels disagreed typically by 20
percent. The mismatches are probably due to variations in guiding accuracy
and in the seeing (which usually was comparable to the slit width) during
the observations, as well as to inaccuracies in the IR magnitudes of the
calibration stars used in the reduction.  Another possibility is
variations of U Equ itself. However, the two 3-4~$\mu$m segments, which
were obtained more than two years apart, were mismatched by only 20\%. The
spectral segments were scaled by small factors in order to adjoin them
smoothly. The final flux calibrations are believed to be accurate to
+/-20\%.

\section{Results}

\subsection{Overview of 1-4~$\mu$m Spectrum}

Figure 1 shows the medium resolution 1.0--2.5~$\mu$m spectrum of U Equ as
observed in 1997 together with the 2.9--4.1~$\mu$m spectrum as observed in
1998 and 2000. The 1.4--2.5~$\mu$m spectrum obtained in 2003 is closely
similar, in depths of absorption bands and in overall continuum slope, to
the spectrum in the top panel of Fig.~1. The 1-4~$\mu$m spectrum is
dominated by water vapor absorption, which is responsible for the deep
bands at 1.1--1.2~$\mu$m, 1.3--1.5~$\mu$m and 1.8--2.0~$\mu$m as well as
the detailed spectral structure at 2.9--3.4~$\mu$m. Weak spectral
structure is also evident longward of 3.7~$\mu$m (see Fig.~1, lower
panel), but we are unable to identify the absorbing species at this
spectral resolution and signal-to-noise ratio ($\sim$50). Also apparent in
the spectrum is absorption due to the first overtone band of carbon
monoxide at 2.3--2.4~$\mu$m. The 1.8--2.0~$\mu$m water band had been seen
previously in a 1.6-2.5~$\mu$m survey spectrum by Lancon \& Wood (2000),
but the CO absorption was not reported by those authors, although in
retrospect it appears to be present in their spectrum at a low
signal-to-noise ratio.

The shape of the 2.3--2.4~$\mu$m spectrum in Fig.~1, with its local
maximum at the CO 2-0 band center near 2.343~$\mu$m, indicates that most
of the CO that is in the lowest (v=0) vibrational state. Thus, the CO is
much cooler than a stellar photosphere and must reside in circumstellar
material. The narrowness of the 1.9~$\mu$m water band in comparison to
those of cool Miras (e.g., see Lancon \& Wood 2000) also suggests that the
absorbing material is quite cool.

\subsection{Higher resolution spectra: CO, H$_{2}$O, and H$_{2}$}

A higher resolution (R$\sim$4100) spectrum of the CO first overtone band
obtained in 1998 is shown in Fig. 2. In it the vibration-rotation
structure of the P and R branches 2-0 band of CO is clearly resolved into
lines. The strongest lines correspond to J$\sim$10, but individual lines
in the R branch can be seen out to J=25 at this resolution. As in the
medium resolution spectrum, no 2-0 band head (corresponding to J$\sim$50)
is apparent at 2.293~$\mu$m. The unevenness of the band structure
indicates that absorbers and possibly emitters other than CO are present
as well, especially near 2.320-2.325 and 2.340--2.345~$\mu$m. $^{13}$CO
and other rarer isotopic species of CO cannot be significant contributors
in either of these spectral intervals.

Fig. 3 shows the portion of the R=4100 spectrum from 1998 in more detail,
as well as the same interval observed with the CGS4 at R=18000 in 2003.  
This region contain the strongest CO lines.  The difference in the
observed widths and depths of the lines is consistent with the different
resolutions. From the 2003 spectrum we derive a deconvolved FWHM of
35~km~s$^{-1}$ for these lines. At both times the lines were centered at
the nominal radial velocity, -91~km~s$^{-1}$ hel. At the higher resolution
of the 2003 spectrum several weaker lines are present. With the aid of the
HITRAN2004 database (Rothman et al. 2003) we have identified them as
H${_2}$O from relatively low-lying energy levels.

Fig. 4 is an echelle spectrum in the vicinity of the CO 2-0 band head. As
was apparent in the lower resolution spectra of Figs. 1 and 2, the band
head is not seen.  The highest J levels that can be distinguished are near
J=30. H$_{2}$O accounts for most of the strong absorption lines in this
spectral interval.

The spectrum in Fig. 5 covers an interval near 2.38~$\mu$m containing the
strongest lines of the 2-0 R branch of $^{13}$CO.  Most of the $^{13}$CO
lines are blended with stronger lines of H$_{2}$O and/or the $^{12}$CO 2-0
P branch and are not readily apparent. However, the R(10) and R(11) lines
of $^{13}$CO are fairly well isolated and clearly detected and the R(12)
line also is apparent between lines of the other species. Equivalent
widths cannot be determined with accuracy because of uncertainties both in
the amounts of line blending and in the placement of the continuum.  

A high resolution spectrum also was obtained at the wavelength of the 1-0
S(1) transition of H$_{2}$ and is shown in Fig.~6.  This line was detected
and its equivalent width over a 75~km~s$^{-1}$ interval centered at the
stellar velocity measured to be 9$\pm$2~$\times$~10$^{-6}$~$\mu$m. The
core of the line appears to be noticeably narrower than the CO, but this
could easily be a consequence of the low signal-to-noise ratio of the
spectrum.

\section{Discussion}

\subsection{CO Temperature, $^{12}$C/$^{13}$C, and Column Densities}

We have attempted to fit the $^{12}$CO absorption spectrum in Fig.~2 with
a multilayer slab models. Spectra of several single temperature slabs are
shown in Fig.~7. Even though other absorbing species, principally
H$_{2}$O, contaminate the spectrum it is clear that at temperatures
between 500~K and 1000~K this simple model gives a good general
representation of the CO absorption.  If the central star is a mid-G giant
as suggested by Barnbaum et al. (1996), and all of the CO is at 600~K, the
CO is located roughly 5~$\times$~10$^{13}$~cm (3 a.u.) from the star. At a
distance of 1.5~kpc, a disk or shell of this dimension subtends only
0.005$\arcsec$ on the sky and would not be resolved by adaptive optics on
the largest telescopes.  Indeed upper limits to the diameter from HST
imaging (C. Barnbaum 1996, unpublished data) and Keck adaptive optics
imaging (M. Morris 2004, unpublished data) are each $\sim$0.05$\arcsec$.

Figure 8 shows four-layer (2000K, 1500K, 1000K, and 500K) model spectra
near 2.38~$\mu$m, the same interval shown in Fig.~5 containing isolated
lines of $^{13}$CO and numerous lines of H$_{2}$O. Each synthetic spectrum
has a $^{12}$CO column density of 2~$\times$~10$^{20}$~cm$^{-2}$, which
fits the $^{12}$CO line strengths well. The upper panel shows the effect
of varying the H$_{2}$O column density for $^{12}$C/$^{13}$C~=~4, whereas
the lower panel shows the effect of varying $^{12}$C/$^{13}$C. As noted
previously many of the CO lines are blended with H$_{2}$O lines; in
addition either the H$_{2}$O line list is incomplete or there are other
contaminants.  Nevertheless, adding H$_{2}$O noticeably improves the fit,
and these simple models suggest that $^{12}$C/$^{13}$C $\sim$4 and that
H$_{2}$O is approximately 2.5 times more abundant than CO.

For the above CO column density and temperature, and at the observed line
width, the strongest lines of $^{12}$CO have optical depths at line center
of order unity in the lowest temperature layer, which contains roughly
half of the CO. Therefore the best fit model spectrum with
$^{12}$C/$^{13}$C~=~4 probably only moderately underestimates
$^{12}$C/$^{13}$C, and we conclude that the ratio is $\leq$10. Assuming
U~Equ has [C]/[H] = 2.6~$\times$~10$^{-4}$ (Asplund 2003) and that all
carbon is in CO, the measured CO column density implies
N(H$_{2}$)~$\sim$4~$\times$10$^{23}$~cm$^{-2}$.  Using the formulae in
Lacy et al. (1994) the measured equivalent width of the H$_{2}$ S(1) line
yields an H$_{2}$ column density of 4~$\times$~10$^{23}$~cm$^{-2}$ in the
v=0, J=1 state, which is roughly half the total column density of H$_{2}$
at temperatures of 500--1000~K. The comparison suggests that gas phase
carbon might be depleted by a factor of roughly 2.

The column densities of CO and H$_{2}$ in U Equ's circumstellar material
are remarkably high. Justannont et al. (1996) found a CO column density of
1~$\times$~10$^{19}$~cm$^{-2}$ in the circumstellar gas of the oxygen-rich
mass-losing supergiant NML~Cygni and Hall \& Ridgway (1978) found
1~$\times$~10$^{19}$~cm$^{-2}$ of CO in the mass-losing carbon star
IRC+10216. The optical extinctions to these two evolved stars are $\sim$15
and $\sim$20 magnitudes, respectively (Ridgway et al. 1986; Ivesi\'c \&
Elitzur 1996), much greater than the 2.1 mag toward U Equ (Barnbaum et al.
1996). Column densities of warm CO observed in the dense winds of young
stellar objects (e.g., Mitchell et al. 1990) typically are more than two
orders of magnitude less than toward U Equ. In interstellar material a gas
column density such as that observed in U Equ would be associated with a
visual extinction of several tens of magnitudes as well as an H$_{2}$
column density of $\sim$10$^{23}$~cm$^{-2}$ (using the cosmic value of
[C]/[H]). The former is in conflict with the estimate of Barnbaum et al.
(1996) and is clearly unrealistic in view of the visual brightness of U
Equ.

Thus, it appears that either the gas-to-dust ratio is much higher in U Equ
than in typical circumstellar or interstellar matter, or the dust and
associated gas are located in a thin disk that transmits or scatters
(above and below the disk) a significant fraction of the photospheric
light. In the latter case the above column density is a lower limit. 

\subsection{Velocities}

The CO line profiles, which are centered on the stellar velocity and are
symmetric about it, show no evidence for net infall or outflow of the
circumstellar material. This result is consistent with the OH line
measurements by Sivagnaman et al. (1990), which indicate radial motions of
only a few km~s$^{-1}$.  A gaseous shell at T$\sim$600~K stably orbiting U
Equ at radius $\sim$5~$\times$~10$^{13}$~cm, and observed in absorption
against a central star, would produce much narrower absorption lines than
the observed 35~km~s$^{-1}$ FWHM. If the gas is in a shell-like structure,
its motion is highly turbulent and has maintained that turbulence over 6
years, which does not seem likely.

The orbital speed of material at the inferred distance of
$\sim$5~$\times$~10$^{13}$~cm from a roughly solar mass star is
20~km~s$^{-1}$. Thus, the observed 35~km~s$^{-1}$ linewidths suggest that
the infrared continuum could be highly extended and nearly coincident with
the absorbing gas. The continuum would then result from scattering,
absorption, and reemission of photospheric radiation. This crude model
probably can account for the observed broad lines but cannot explain the
low extinction.

\section{Conclusion}

Like the optical spectrum, the infrared spectrum of U Equ appears to be
unique. Its most remarkable qualities are the presence of large amounts of
warm circumstellar CO and water vapor and, as surmised from the optical
spectrum, an extreme lack of obscuring dust. The abundances in U Equ are
not well quantified, but certainly the presence in the optical spectrum of
strong TiO and other metallic oxide features suggests that the ingredients
for forming copious amounts of dust are present.  The temperature of the
CO is well below the dust condensation temperature of metal oxides.  Thus
all evidence to date suggests that U Equ should be surrounded by a totally
obscurring cloud of dust. That this is not the case probably either
implies a remarkably high gas to dust ratio or peculiar geometry and
illumination, perhaps with starlight illuminating and heating the edge of
a very thin circumstellar disk-like structure.

It has been proposed that the phenomena being seen in U Equ might be
consistent with the destruction of a planet. Siess \& Livio (1999)
suggested that U Equ might be extruding metal- enriched material because
of rapid rotation caused by the accretion of a giant planet. However,
there is no evidence for ejection (i.e., no net outflow of gas; indeed not
even asymmetry in the line profiles). Moreover, the very low value of
$^{12}$C/$^{13}$C in the circumstellar material indicates that material is
not highly diluted by planetary matter, which would have a high value of
$^{12}$C/$^{13}$C. Mass estimates are very uncertain because of the
unknown geometry of the absorbing CO. If [C]/[H] is solar and even if the
observed CO is in a spherical shell of radius 5~$\times$~10$^{13}$~cm, the
mass of the shell is only 0.03 that of Jupiter, which, when combined with
the observed value of $^{12}$C/$^{13}$C again suggests that there has been
very little dilution by a planet.

Although the current study provides information on the evolutionary state
and some of the general properties of its circumstellar material, it has
not solved the enigma that is U Equ. In particular, the tools of very high
resolution spectra, polarimetry, and interferometry should now be brought
to bear on this most fascinating object.

\begin{acknowledgements}

We wish to thank the staff of the Joint Astronomy Centre for its support
of these observations.  UKIRT is operated by the Joint Astronomy Centre on
behalf of the U.K. Particle Physics and Astronomy Research Council.  
TRG's research is supported by the Gemini Observatory, which is operated
by the Association of Universities for Research in Astronomy, Inc., on
behalf of the international Gemini parthership of Argentina, Australia,
Brazil, Canada, Chile, the United Kingdom and the United States of
America.

\end{acknowledgements}

\vfill\eject

\begin{figure}
\epsscale{1.0}
\plotone{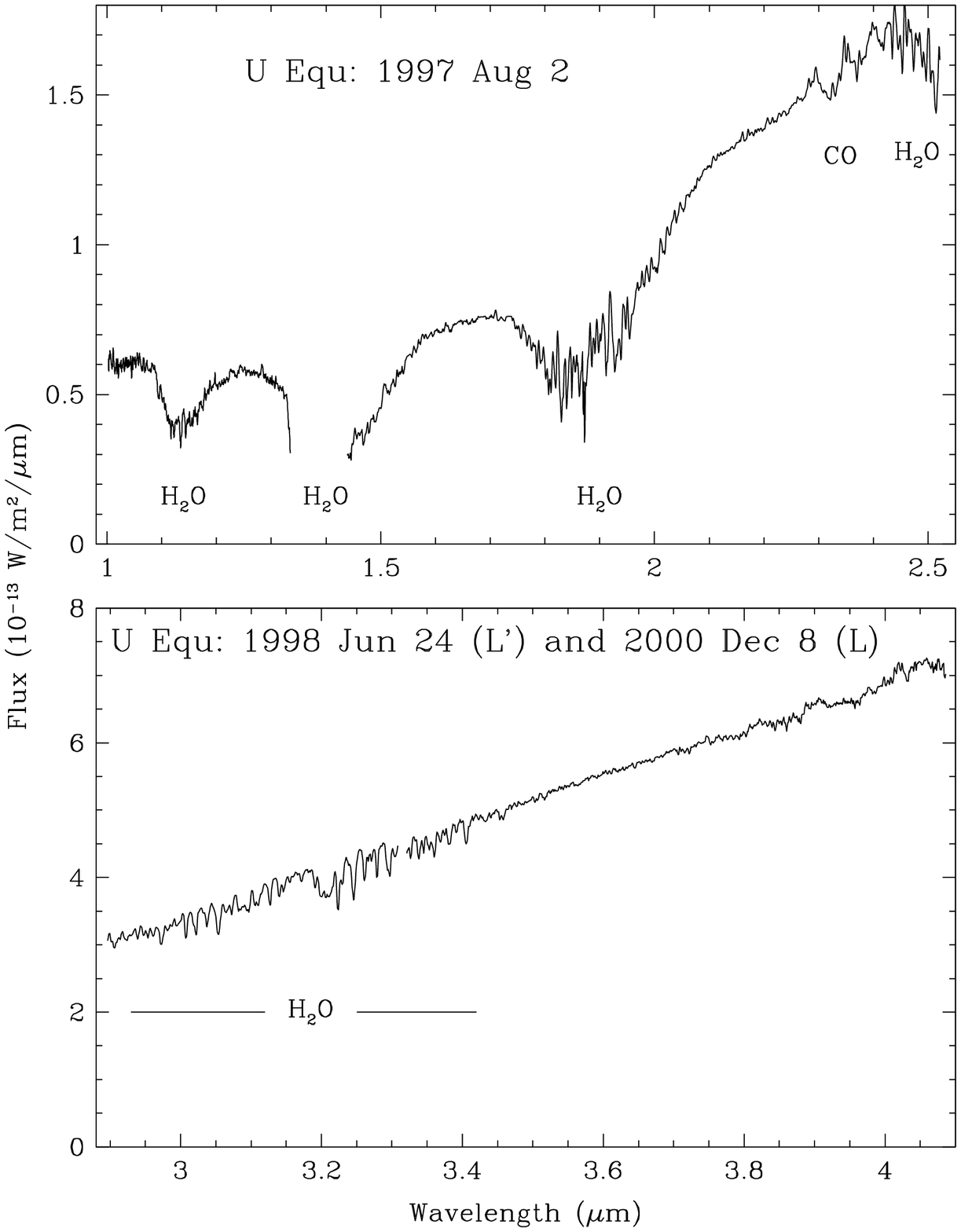}
\caption{Medium resolution 1--4~$\mu$m spectrum of U Equ.
Prominent absorption bands are identified.}
\label{f1}
\end{figure}

\begin{figure}
\epsscale{1.0}
\plotone{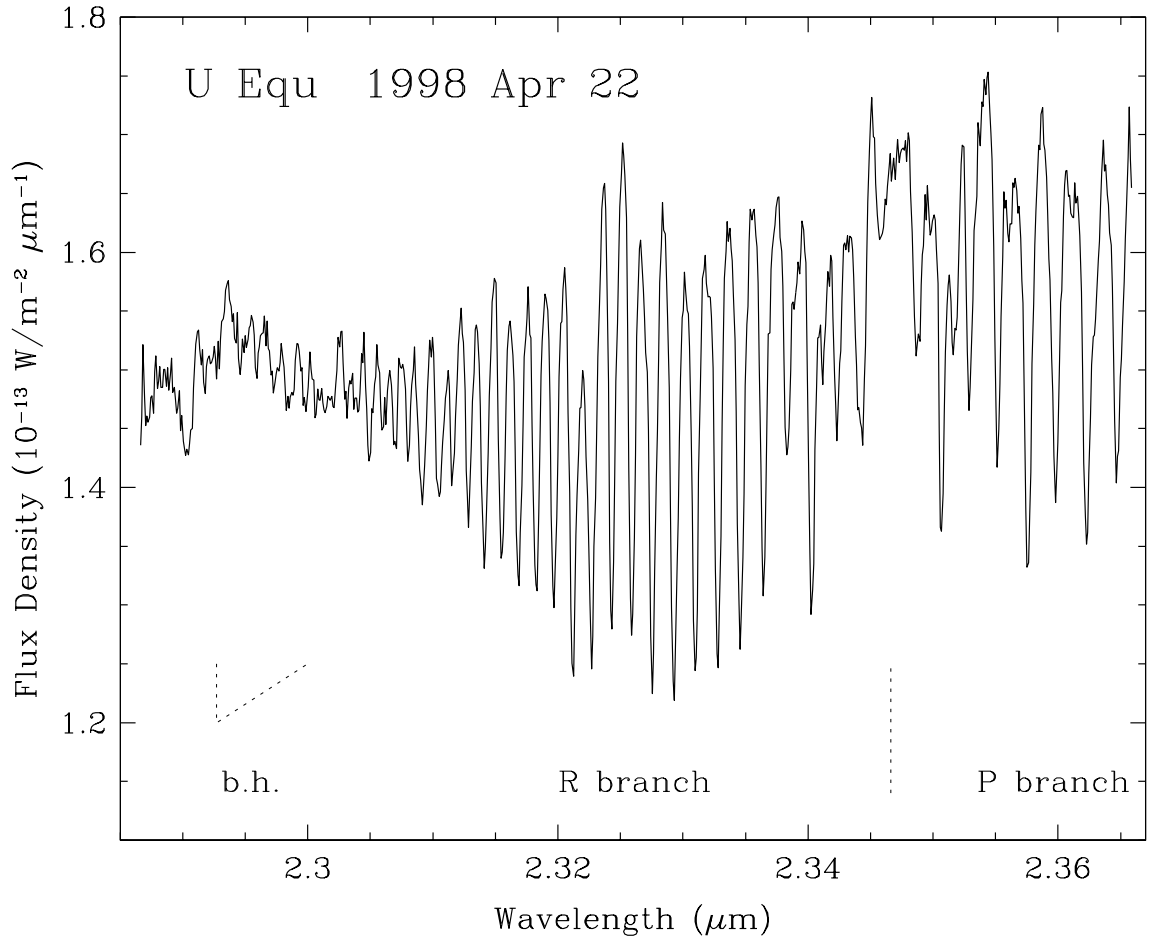}
\caption{Spectrum of U Equ in the 2--0 band of CO. The P and R
branches, band center, and the unseen band head at 2.2927~$\mu$m are
indicated.}
\label{f2}  
\end{figure}

\begin{figure}
\epsscale{1.0}
\plotone{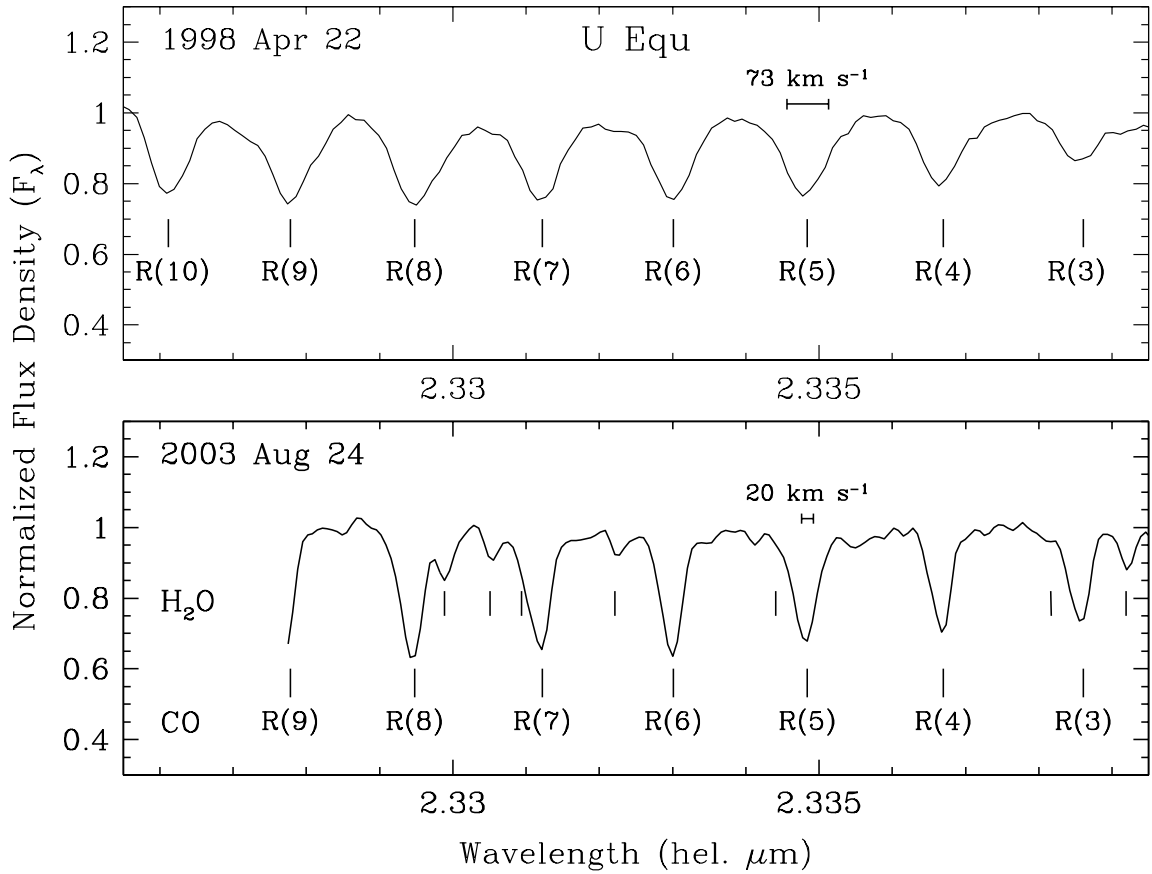}
\caption{Details of the CO 2-0 R branch lines as observed in 1998
and 2003. The wavelength scale is for zero heliocentric velocity. Lines of
CO and known strong lines of H$_{2}$O are identified; their positions,
indicated by vertical lines, are shifted by -91~km~s$^{-1}$, the
heliocentric velocity of U Equ. Velocity resolutions are also shown.}
\label{f3}
\end{figure}

\begin{figure}
\epsscale{1.0}
\plotone{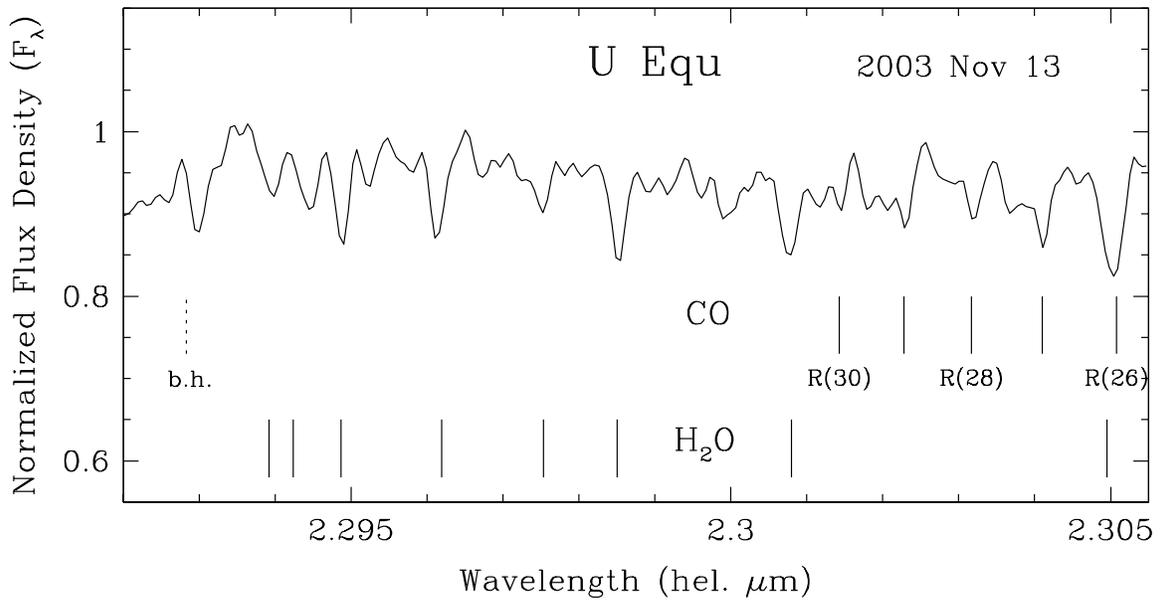}
\caption{High resolution spectrum near the CO 2-0 band head.
Positions of identified lines of CO and H$_{2}$O are shifted by
-91~km~s$^{-1}$. The location of the band head is also shown.}
\label{f4}
\end{figure}

\begin{figure}
\epsscale{1.0}
\plotone{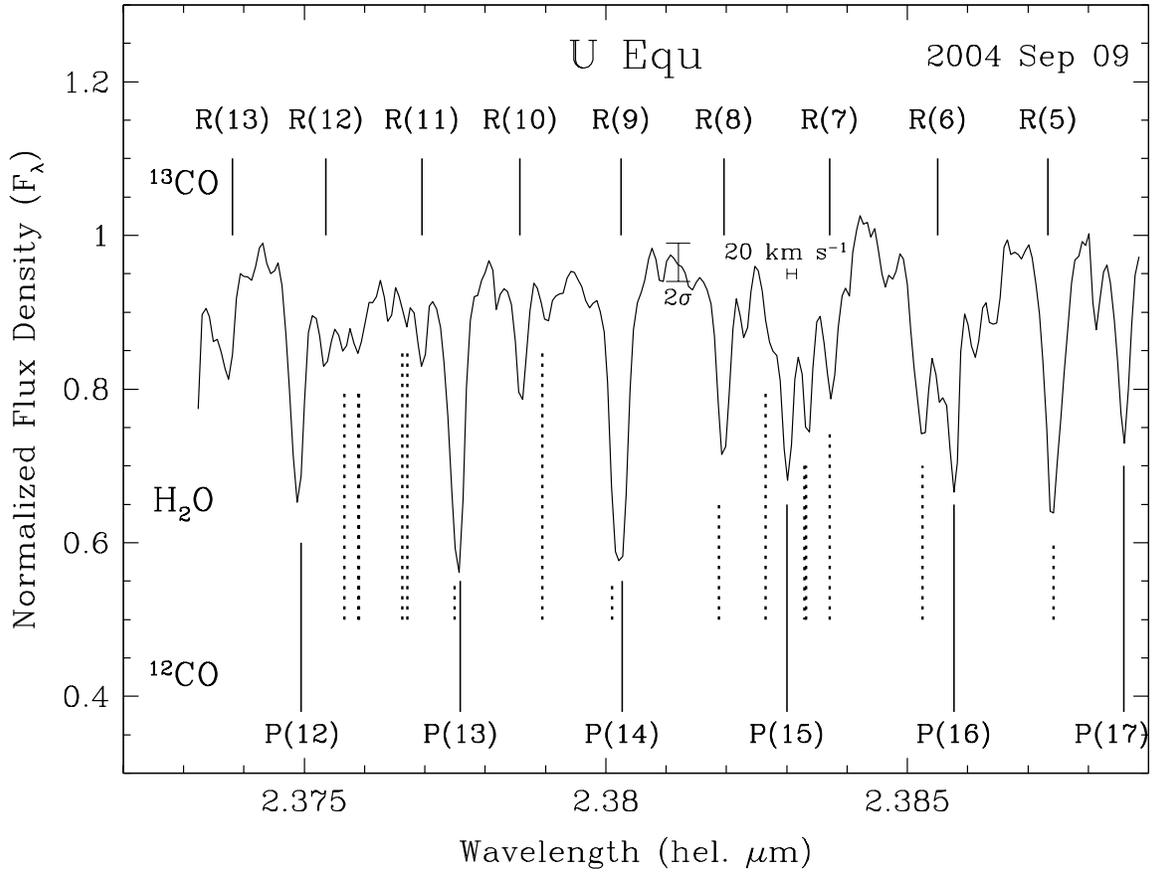}
\caption{High resolution spectrum in the $^{12}$CO 2-0 P branch,
showing detection of R(10) - R(13) lines of $^{13}$CO. All line indicators
are shifted by -91~km~s$^{-1}$; known strong water lines are indicated by
dashed lines.}
\label{f5}
\end{figure}

\begin{figure}
\epsscale{1.0}
\plotone{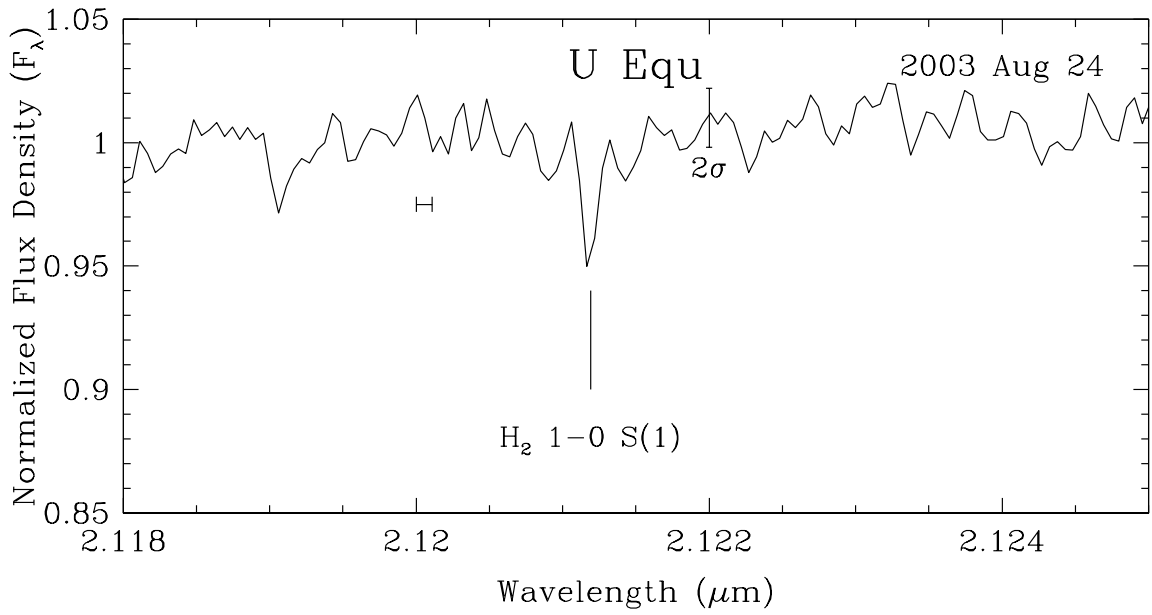}
\caption{Spectrum at the wavelength of the H$_{2}$ 1-0 S(1) line.
The noise level, spectral resolution, and line wavelength, Doppler-shifted
by -91~km~s$^{-1}$ are indicated.}
\label{f6}
\end{figure}

\begin{figure}
\epsscale{1.0}
\plotone{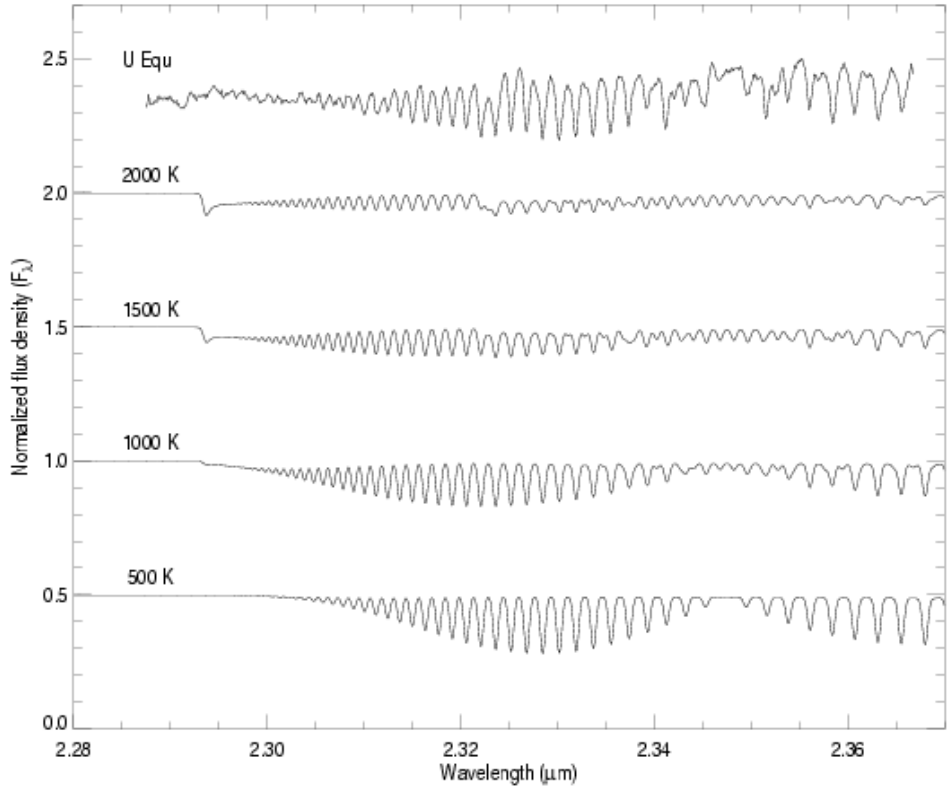}
\caption{The observed R=4000 spectrum of U Equ in the CO 2-0 band
(top) compared with isothermal slab absorption models of a CO column
density of 1~$\times$10$^{20}$~cm$^{-2}$ at four temperatures.}
\label{f7}
\end{figure}

\begin{figure}
\epsscale{1.0}
\plotone{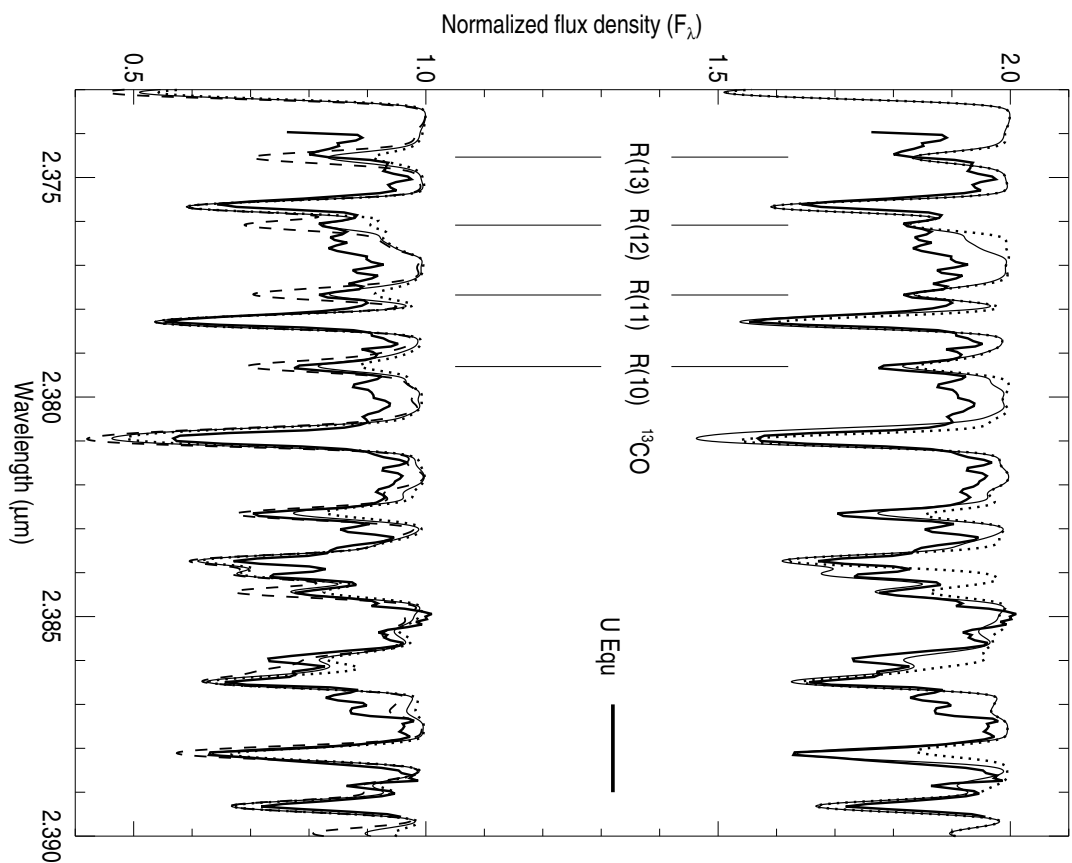}
\caption{The observed R=19000 spectrum of U Equ near 2.38~$\mu$m
(thick line) compared with multislab synthetic spectra with
N($^{12}$CO)~=~2~$\times$~10$^{20}$~cm$^{-2}$ and various amounts of
$^{13}$CO and H$_{2}$). Most of the strong lines are due to $^{12}$CO.
Relatively unblended lines of $^{13}$CO are indicated. The upper panel
shows spectra with $^{12}$CO/$^{13}$C~=~4 and [H$_{2}$O]/[CO]~=~0 (dotted
line) and 2.5 (thin line). The lower panel shows spectra with
$^{12}$C/$^{13}$C~=~8 (dotted), 4 (continuous), and 2 (dashed), each with
[H$_{2}$O]/[CO]~=~2.5.}
\label{f8}
\end{figure}

\begin{deluxetable}{ccccc}
\tablecolumns{5}
\tablewidth{0pt}
\tablecaption{Observing Log}
\tablehead{
\colhead{Date} &
\colhead{Wavelength ($\mu$m)} &
\colhead{R} &
\colhead{Exposure (s)} &
\colhead{Calib. star}
}
\startdata
19970802 & 1.00-1.33   &   950 &  288 &HR 7973 (F5V) \\
19970802 & 1.44-2.10   &   700 &   96 & HR 7973 (F5V) \\
19970802 & 1.87-2.52   &   880 &   96 & HR 7973 (F5V) \\
19980407 & 2.25-2.42   &  2100 & 1080 & HR 7973 (F5V) \\
19980422 & 2.29-2.37   &  4100 & 1080 & HR 8205 (A1V) \\ 
19980624 & 3.45-4.09   &  1500 &  240 & HR 8328 (A1V) \\
20001208 & 2.89-3.52   &  1280 &  576 & HR 8205 (A1V) \\
20030824 & 1.40-2.50   &   450 & 1800 & HIP 102805 (F5V) \\
20030824 & 2.328-2.344 & 18000 &  960 & HR 8178 (A3V) \\
20030824 & 2.115-2.127 & 21000 & 1920 & HIP 102631 (A0V) \\
20030909 & 2.374-2.388 & 19000 &  480 & HIP 102631 (A0V) \\
20031113 & 2.292-2.305 & 21000 & 1440 & HIP 102631 (A0V) \\
\enddata
\end{deluxetable}

\end{document}